\documentclass[aps,prl,twocolumn,showpacs,amsmath,amssymb,superscriptaddress,10pt]{revtex4-2}

\usepackage[pdftex]{graphicx}
\usepackage[colorlinks=true,citecolor=blue]{hyperref}
\usepackage{braket}
\usepackage{lipsum}

\usepackage{xcolor}
\newcommand{\revision}[1]{{{#1}}}
\newcommand{\revisiontwo}[1]{{{#1}}}
\newcommand{\revisionthree}[1]{{{#1}}}

\newcommand{\tx}{\text}
\newcommand{\fig}[1]{Fig.~\ref{fig:#1}}
\newcommand{\eq}[1]{Eq.~(\ref{eq:#1})}

\renewcommand{\vec}[1]{\ensuremath{\mathbf{#1}}}

\begin{document}
\title{\revision{Electron Waiting Times in a Strongly Interacting Quantum Dot:\\ Interaction Effects and Higher-Order Tunneling Processes}}

\author{Philipp Stegmann}\email{philipp.stegmann@uni-due.de}
\affiliation{Theoretische Physik, Universit\"at Duisburg-Essen and CENIDE, 47048 Duisburg, Germany}

\author{Bj\"orn Sothmann}
\affiliation{Theoretische Physik, Universit\"at Duisburg-Essen and CENIDE, 47048 Duisburg, Germany}

\author{J\"urgen K\"onig}
\affiliation{Theoretische Physik, Universit\"at Duisburg-Essen and CENIDE, 47048 Duisburg, Germany}

\author{Christian Flindt}
\affiliation{Department of Applied Physics, Aalto University, 00076 Aalto, Finland}

\date{\today}

\begin{abstract}
\revision{Distributions of electron waiting times have been measured in several recent experiments and have been shown to provide complementary information compared to what can be learned from the electric current fluctuations. Existing theories, however, are restricted to either weakly coupled nanostructures or phase-coherent transport in mesoscopic conductors. Here, we consider an interacting quantum dot and develop a real-time diagrammatic theory of waiting time distributions that can treat the interesting regime, in which both interaction effects and higher-order tunneling processes are important. Specifically, we find that our quantum-mechanical theory captures higher-order tunneling processes at low temperatures, which are not included in a classical description, and which dramatically affect the waiting times by allowing fast tunneling processes inside the Coulomb blockade region. Our work paves the way for systematic investigations of temporal fluctuations in interacting quantum systems, for example close to a Kondo resonance or in a Luttinger liquid.}
\end{abstract}

\maketitle

\textit{Introduction.}--- Electronic waiting time distributions are an important concept in the analysis of quantum transport in nano-scale structures~\cite{brandes_waiting_2008,Welack_2009,albert_distribution_2011,albert_electron_2012,rajabi_waiting_2013,thomas_electron_2013,dasenbrook_floquet_2014,Albert_2014,haack_2014,sothmann_electronic_2014,tang_waiting_2014,Thomas_2014,Dasenbrook_2015,soutot_transient_2015,chevallier_probing_2016,ALBERT_2016,Emary_2016,Dasenbrook_2016,HOFER_2016,Rudge2016,Kosov_2017,Potanina_2017,ptaszy_waiting_2017,ptaszy_nonrenewal_2017,ptaszy_first_2018,walldorf_electron_2018,tang_spin_2018,mi_electron_2018,Kosov2018,rudge_distribution_2018,rudge_nonrenewal_2019,engelhardt_tuning_2019,nathan_counting_2019,Burset_2019,rudge_counting_2019,wrzeniewski2020}\revision{, and recently they have been measured in several experiments \cite{gorman_tunneling_2017,jenei_2019,matsuo_2019,kurzmann_optical_2019,Brange2021,Ranni2020}}. Unlike full counting statistics, which typically considers the time-integrated current fluctuations~\cite{levitov_electron_1996,bagrets_full_2003,bagrets2006}, waiting time distributions are concerned with the short time-span that passes between the charge transfers in a nanoscale conductor. The  waiting time distribution contains a wealth of information about the underlying transport processes. For example, it has been predicted that the waiting time distribution for a quantum point contact should exhibit a cross-over from Wigner-Dyson statistics at full transmission to Poisson statistics close to pinch-off~\cite{albert_electron_2012,haack_2014}, illustrating a profound connection between free fermions and the eigenvalues of random matrices~\cite{Metha_2004}. 

\revision{Experimentally, waiting time distributions have been used to demonstrate accurate control of the emission time statistics of a dynamic single-electron transistor \cite{Potanina_2017,Brange2021} and to characterize the splitting of Cooper pairs in the time domain \cite{walldorf_electron_2018,wrzeniewski2020,Ranni2020}. Theoretically, electron waiting time distributions have been considered in two opposite regimes. For sequential tunneling in nanostructures, the waiting time distribution can be obtained from a master equation description of the charge transport~\cite{brandes_waiting_2008,albert_distribution_2011,rudge_counting_2019}. On the other hand,} for phase-coherent transport of non-interacting electrons, the waiting time distribution can be evaluated using scattering theory~\cite{albert_electron_2012, haack_2014,dasenbrook_floquet_2014}, tight-binding calculations \cite{Thomas_2014}, or Green's function methods~\cite{tang_waiting_2014,soutot_transient_2015}. However, for the interesting intermediate regime, where both interaction effects and higher-order tunneling processes are important, a systematic theory of electronic waiting time distributions has so far been lacking.

\begin{figure}
	\includegraphics[width=\columnwidth]{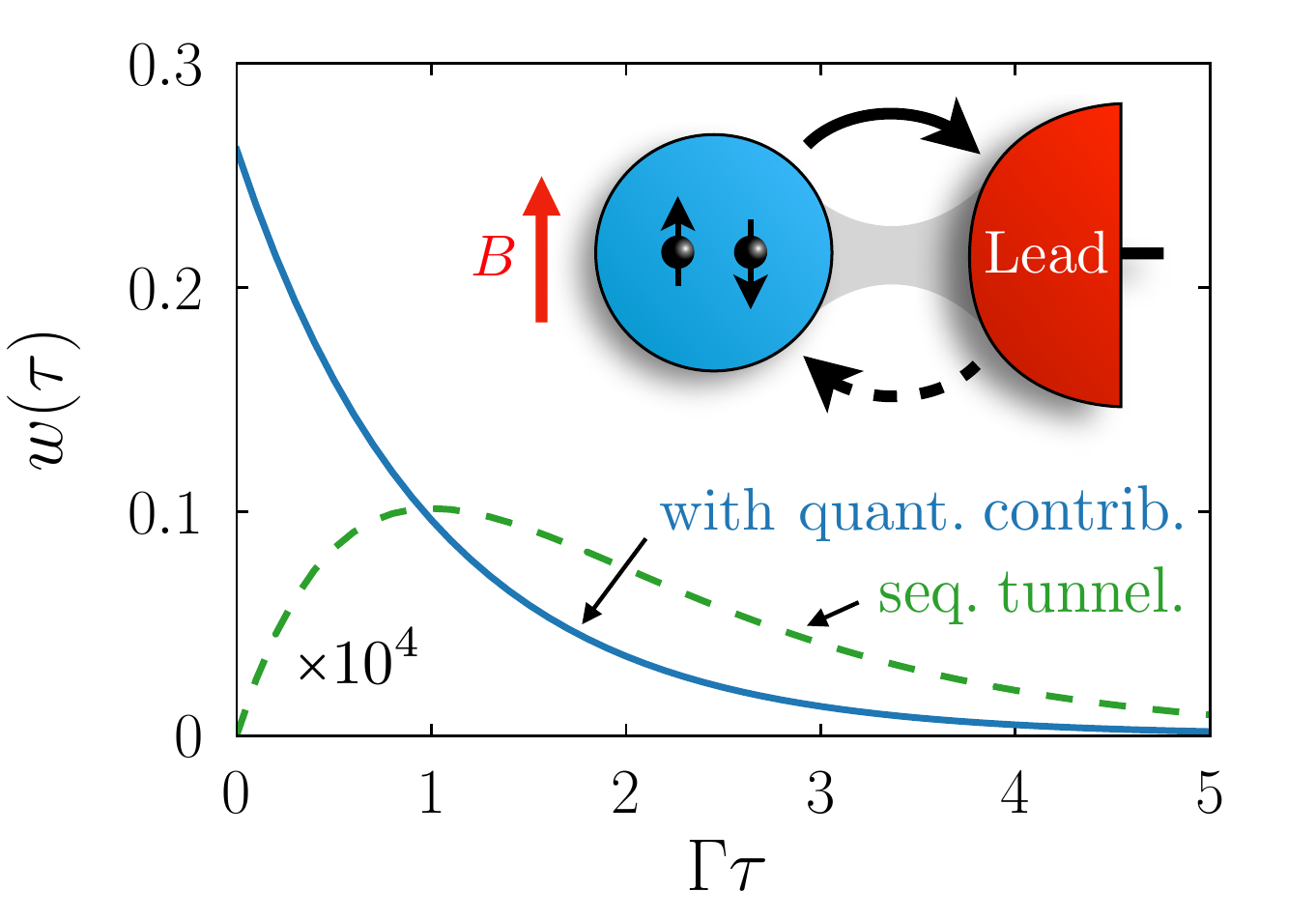}
	\caption{\revision{Electron waiting times for an interacting quantum dot in a magnetic field. Electrons tunnel back and forth (arrows) between the quantum dot and an external reservoir. We consider the waiting time between electrons leaving the quantum dot (full arrow). Waiting time distributions are shown for lowest-order sequential tunneling only (green curve) as well as with higher-order processes included (blue curve). The level position with respect to the chemical potential of the lead is $\varepsilon=-0.2U$, where $U$ is the Coulomb interaction, and the magnetic field causes the Zeeman splitting $\Delta =1.1U$. The temperature and the coupling \revisiontwo{are $k_\text{B}T=U/30$, $\hbar\Gamma=10^{-7} U$.}}
	}\label{fig:fig1}
\end{figure}

In this Letter, \revision{we investigate the temporal fluctuations of charge tunneling in a Coulomb-blockade quantum dot and develop a real-time diagrammatic theory of electron waiting time distributions that can treat both strong interactions and higher-order tunneling processes~\cite{koenig_diagram_1996,timm_tunneling_2008,koller_density_2010}. Cotunneling processes can be described at the level of mean currents using $T$-matrix approaches~\cite{timm_tunneling_2008,koller_density_2010, kaasbjerg_full_2015}, which, however, cannot account for non-Markovian effects that influence the charge transport fluctuations beyond average values~\cite{braggio_full_2006, flindt_counting_2008}. Our setup is illustrated in Fig.~\ref{fig:fig1}, where we also show waiting time  distributions for sequential tunneling only as well as with  higher-order tunneling processes included. As we will see, the higher-order tunneling processes are predominantly of quantum nature, and they tend to get washed out by an increasing electronic temperature for which we recover a classical description based on sequential tunneling. While we here focus on quantum dots, it will become clear that our approach can be applied to a wide range of interacting nanostructures at low temperatures and bias voltages, where quantum effects are important \cite{ridley_numerically_2018,ridley_numerically_2019,Ridley2019_chem,mundinar_iterative_2019,kilgour_path_2019,schinabeck_hierarchical_2020,erpenbeck_revealing_2020}.}

\revision{\textit{Hamiltonian.}--- We start by considering a quantum dot coupled to a single electrode before extending the discussion to a transport setup with both a source and a drain electrode. The full Hamiltonian of the setup reads 
\begin{equation}
\mathcal{\hat H}=\mathcal{\hat H}_\tx{qd}+\mathcal{\hat H}_\tx{res}+\mathcal{\hat H}_\tx{tun},
\end{equation}
where the quantum dot,  $\mathcal{\hat H}_\text{qd}= \sum_{\sigma}\varepsilon_\sigma \hat d^{\dagger}_\sigma \hat d_\sigma +U \hat d^{\dagger}_\uparrow \hat d_\uparrow \hat d^{\dagger}_\downarrow \hat d_\downarrow$, can be either empty $\ket{0}$, occupied by a single electron $\ket{\sigma}$ with spin $\sigma=\uparrow,\downarrow$, or doubly occupied $\ket{\tx{d}}$, with $U$ denoting the onsite Coulomb interactions. An applied magnetic field lifts the degeneracy of spin-up, $\varepsilon_\uparrow =\varepsilon+\frac{\Delta}{2}$, and spin-down electrons, $\varepsilon_\downarrow =\varepsilon-\frac{\Delta}{2}$, where $\Delta$ is the Zeeman splitting. The orbital energy $\varepsilon$ relative to the chemical potential $\mu$ of the reservoir can be tuned by an external gate voltage. The reservoir contains non-interacting electrons and is described by the Hamiltonian $\mathcal{\hat H}_\tx{res}=\sum_{\vec k \sigma} \varepsilon_{\vec k \sigma} \hat a^{\dagger}_{\vec k \sigma} \hat a_{\vec k \sigma}$, while electron tunneling between the quantum dot and the reservoir is governed by the Hamiltonian $\mathcal{\hat H}_{\tx{tun}}=\sum_{ \vec k  \sigma} (g\, \hat a_{\vec k \sigma}^{\dagger} \hat d_{ \sigma} + g^* \, \hat d_{\sigma}^{\dagger}\hat a_{\vec k \sigma})$. The tunnel coupling between the quantum dot and the lead is given by $\Gamma=2\pi | g |^2 \nu/\hbar$, where $\nu$ is the density of states in the lead, which is kept at temperature $T$.

\textit{Master equation.}--- We describe the quantum dot by its density matrix $\hat\rho_N(t)$, which is resolved with respect to the number of transferred particles, so that $P_N(t)=\tx{Tr}[\hat\rho_N(t)]$ is the probability that $N\ge 0$ electrons have tunneled into the reservoir during the time span $[0,t]$~\cite{Plenio1998,Makhlin2001,bagrets_full_2003}. The dynamics of the quantum dot is governed by the non-Markovian master equation~\cite{braggio_full_2006,flindt_counting_2008,flindt_2010,zwanzig_theory_2001}}
\begin{equation}\label{eq:qme}
\frac{d}{d t} \hat\rho_N(t) =  \sum_{N'} \int\limits_{0}^t dt'\, \mathbb{W}_{N-N'}(t-t') \hat\rho_{N'}(t')+\hat\gamma_N(t)\, ,
\end{equation}
\revisiontwo{where the kernel $\mathbb{W}$ is obtained by evaluating real-time diagrams on the Keldysh contour up to the desired order in the tunnel coupling~$\Gamma$~\cite{koenig_diagram_1996,SM}. The inhomogeneity $
\hat\gamma_N(t) = \int^{0}_{-\infty}dt'\,\widetilde{\mathbb{W}}_{N}(t,t')\hat\rho_\text{stat}$ describes correlations that build up between the quantum dot and the lead before the counting of particles begins at $t=0$~\cite{flindt_counting_2008,flindt_2010,marcos_non_2011,emary_quantum_2011,thomas_electron_2013}, and it is expressed in terms of a modified kernel $\widetilde{\mathbb{W}}$, which extends to earlier times.} The quantum dot evolves from an arbitrary state in the far past and has reached its stationary state $\hat\rho_\text{stat}$ well before counting begins.

Solving the non-Markovian master equation is a formidable task. However, we can use an operator-valued generalization of the standard generating function technique in Laplace space by introducing the transformed density matrix $\hat\rho_s(z)= \int_{0}^\infty dt e^{-z t} \sum_N  s^N  \,\hat\rho_N(t)$ together with similar definitions for the kernel and the inhomogeneity.  
Using these definitions, the solution of \eq{qme} \revision{in Laplace space} then becomes~\cite{flindt_counting_2008,thomas_electron_2013}
\begin{equation}\label{eq:rhos}
\hat\rho_s (z)=  \frac{1}{z-\mathbb{W}_{\! s}(z)}\, [\hat\rho_\tx{stat}+\hat\gamma_s(z)] \, ,
\end{equation}
which is a powerful formal result that in principle yields the full distribution of transferred charge for any observation time. As we now will show, it also leads to a systematic theory of electron waiting times in nano-scale conductors, which can include both interaction effects and higher-order tunneling processes.

\textit{Electron waiting times.}--- The waiting time distribution is the probability density that two consecutive tunneling events are separated by the time~$\tau$~\cite{brandes_waiting_2008}. It can be expressed as $w(\tau)=\braket{\tau} \partial^2_\tau\Pi(\tau)$, where $\braket{\tau}=-1/[\partial_\tau\Pi(0)]$ is the mean waiting time, and $\Pi(\tau)=P_{N=0}(\tau)$ is the idle-time probability that no transfers have occurred during the time span $[0,\tau]$~\cite{albert_electron_2012,haack_2014}. Importantly, the idle-time probability can be obtained from \eq{rhos} using that $\Pi(z)=\tx{Tr}\{\hat\rho_{s=0}(z)\}$. By expanding the kernel around $z=0$, we can return to the time domain and find~\cite{SM}
\begin{equation}\label{eq:wfull}
w(\tau) = \sum_{m=0}^{\infty} \frac{\braket{\tau}}{m!}  \partial_z^m \tx{Tr}\left\{ \mathcal{J}(z) \mathbb{W}_{\! 0}^{m}(z)e^{\mathbb{W}_{\! 0}(z)\tau}  \widetilde{\mathcal{J}}(z)   \hat\rho_\text{stat}\right\}_{\! z=0},
\end{equation}
which is a general result that enables a systematic analysis of how tunneling processes of different orders contribute to the distribution of waiting times. Here, we have defined the jump operators $\mathcal{J}(z)=\mathbb{W}_{\!1}(z)-\mathbb{W}_{\! 0}(z)$ and $\widetilde{\mathcal{J}}(z)=\mathcal{J}(z)+ \int_{0}^\infty dt \, ze^{-z t} \int_{-\infty}^0 dt' \, [\widetilde{\mathbb{W}}_{\! 1}(t,t')-\widetilde{\mathbb{W}}_{\! 0}(t,t')]$, and we note that memory effects are encoded in non-zero derivatives with respect to $z$. Thus, only the lowest order term ($m=0$) remains in the Markovian limit, where we recover the well-known result~\cite{brandes_waiting_2008}
\begin{equation}\label{eq:wMa}
w_\tx{M}(\tau)= \braket{\tau}\text{Tr}\left[ \mathcal{J} e^{\mathbb{W}_{\! 0} \tau}\mathcal{J} \hat\rho_\text{stat}\right],
\end{equation}
since the kernel in that case does not depend on $z$. \revision{Equation~(\ref{eq:wMa}) is useful to evaluate waiting time distributions in the sequential tunneling regime, where the charge transport can be described by a classical rate equation. By contrast, Eq.~(\ref{eq:wfull}) allows us to consider lower temperatures, where quantum effects become important.}

\begin{figure*}
	\includegraphics[width=0.96\textwidth]{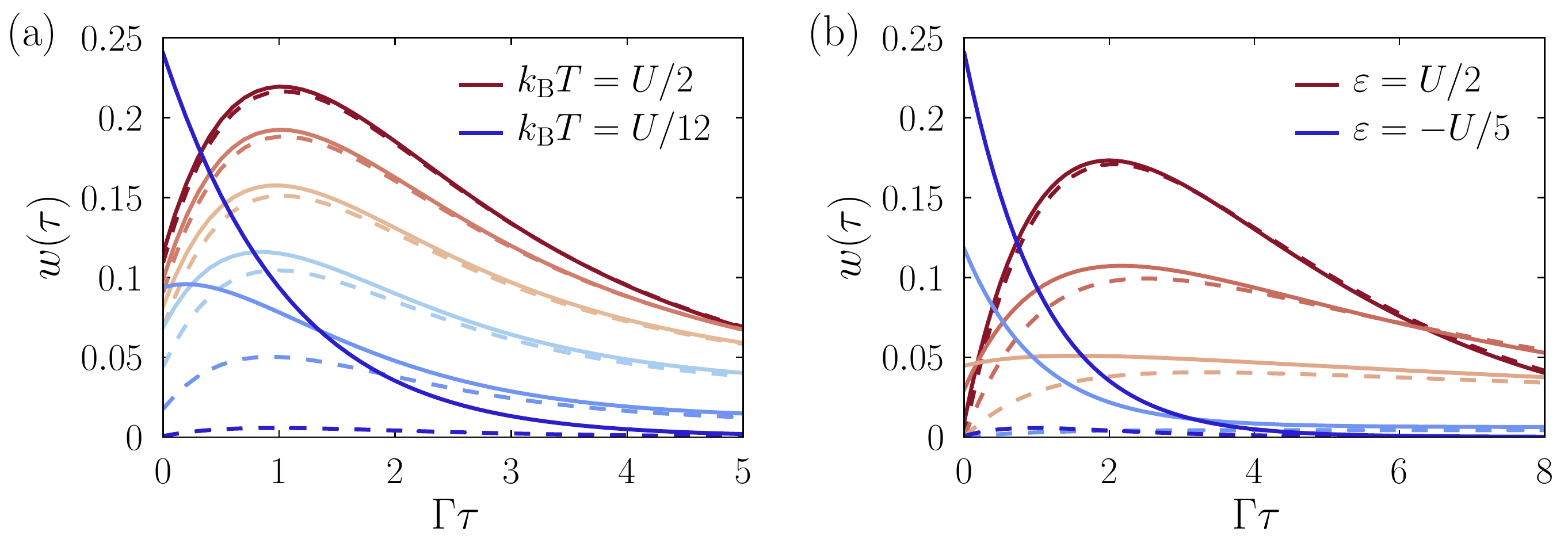}
	\caption{\revision{Distribution of electron waiting times. The quantum mechanical results are based on Eq.~(\ref{eq:wfull}) evaluated up to second order in the tunnel coupling \revisiontwo{$\hbar\Gamma=U/60$} (solid lines), while the results for sequential tunneling using Eq.~(\ref{eq:wMa}) is evaluated up to first order (dashed lines). In panel (a), the temperature is increased in equidistant steps from $k_\text{B}T=U/12$ to  $k_\text{B}T=U/2$, and we use $\varepsilon=-0.2U$ and $\Delta =1.1U$ for the level position and the Zeeman splitting as in \fig{fig1}.
	In panel (b), the level position is $\varepsilon/U=-0.2, 0.1, 0.3, 0.4,$ and $0.5$, and the temperature and Zeeman splitting are $k_\text{B}T=U/12$ and $\Delta =1.1U$.}}\label{fig:fig2}
\end{figure*}

\revision{\textit{Tunneling processes}.---}
To evaluate the waiting times between electrons tunneling out of the quantum dot, we expand the kernel order-by-order in the tunneling coupling $\Gamma$ as $\mathbb{W}_{\! s}(z)=\sum_{n=1}^\infty \,\mathbb{W}_{\! s}^{(n)}(z)$, with similar expressions for the jump operators and the density matrix. Each term is evaluated using the real-time diagrammatic technique for quantum transport in nano-structures~\cite{koenig_diagram_1996,timm_tunneling_2008,koller_density_2010, SM}. Within this framework, we analytically evaluate the waiting time distribution in Eq.~(\ref{eq:wfull}) up to second order in the coupling and compare our results with the expression in Eq.~(\ref{eq:wMa}) using sequential tunneling rates~\cite{SM}. \revisiontwo{The expansion is controlled by the coupling over the interaction, $\hbar\Gamma/U$. In addition, some sequential tunneling rates get exponentially suppressed at low temperatures and are comparable to second-order processes. Higher-order processes, by contrast, can safely be ignored.} 

\revision{Figure \ref{fig:fig2} shows distributions of waiting times between electrons tunneling out of the quantum dot as indicated by the full arrow in Fig.~\ref{fig:fig1}. Our quantum mechanical calculations are presented with solid lines, while results based on sequential tunneling are indicated with dashed lines. In panel (a), we first focus on waiting time distributions for different electronic temperatures. The state of a single spin-down electron is situated well below the chemical potential of the lead, while the state of a spin-up electron is positioned slightly above the chemical potential. The doubly-occupied state has a larger energy.

At low temperatures, we observe marked differences between the quantum mechanical results that include second-order processes and the classical description based on sequential tunneling only. For the classical description, the quantum dot is likely occupied by a spin-down electron, which energetically is situated below the chemical potential of the lead. It is unlikely that the quantum dot is occupied by a spin-up electron or even doubly-occupied. It only rarely happens that the spin-down electron tunnels out of the quantum dot, which is then quickly refilled by a new spin-down electron, which then tunnels out again much later. Thus, there is a long waiting time between electrons tunneling out of the dot as seen in the corresponding waiting time distribution. 

The quantum mechanical results reveal a completely different physical picture with a large peak at short waiting times (solid blue line). The higher-order processes that are now included lead to a renormalized tunnel coupling as well as an increased tunneling rate that makes it possible for the quantum dot to \revisionthree{be doubly-occupied~\cite{SM}}. From the doubly-occupied state there are two decay paths with very different dynamics. If the spin-up electron tunnels out first, the quantum dot is left with a spin-down electron, and there will be a long waiting time until this electron leaves the quantum dot. On the other hand, if the spin-down electron first tunnels out, the quantum dot is left in an excited state with the spin-up electron, which then quickly leaves the quantum dot, giving rise to the large peak at short waiting times. Thus, higher-order processes lift the blockade of the tunneling events that occurs when only sequential tunneling is considered.

To corroborate this physical picture, it is instructive to investigate the effects of an increased electronic temperature, where the doubly-occupied state comes into play because of thermal excitations. In this case, the  sequential tunneling processes become dominant, and the classical and quantum descriptions in panel (a) coincide as one would expect at high temperatures. A similar behavior is observed in panel (b), where we gradually shift the level upwards, so that the energy of a spin-down electron comes close to the chemical potential, and the blockade is lifted. In this case, the spin-up state and the doubly-occupied state are energetically out of reach, and the classical and quantum mechanical results again agree.}

\begin{figure*}
	\includegraphics[width=0.96\textwidth]{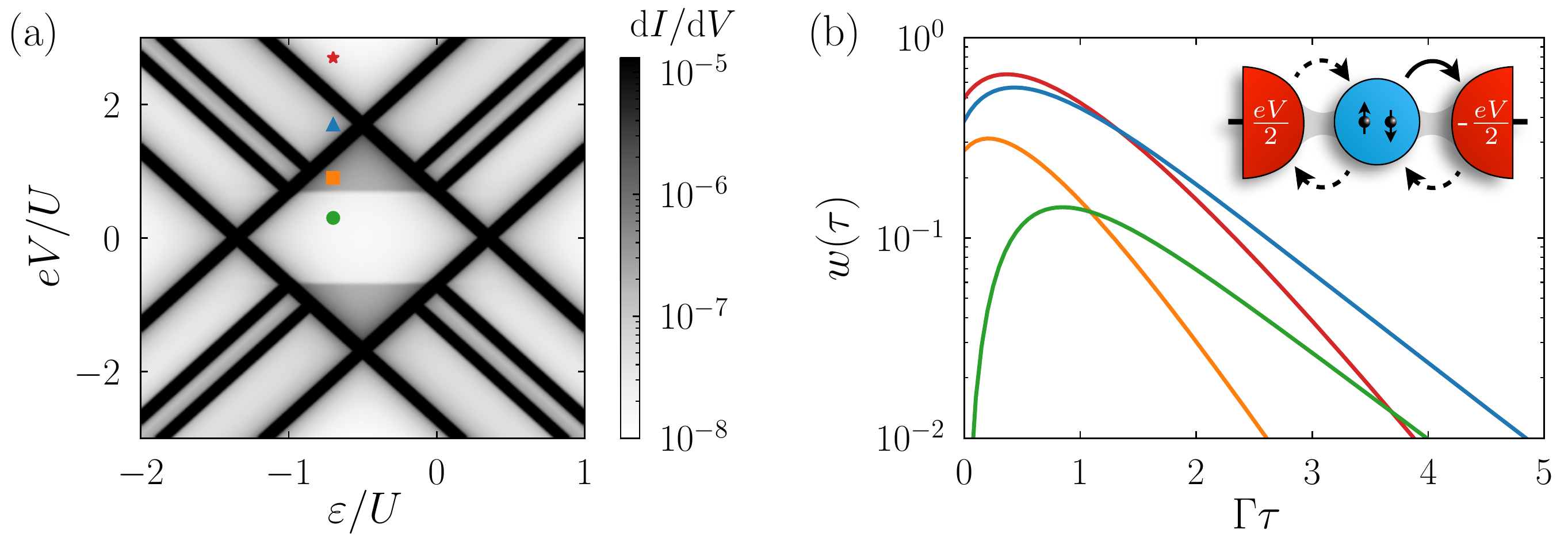}
	\caption{Waiting time distributions for a transport setup. A bias between two electrodes drives a current through the quantum dot. (a) Differential conductance as a function of the bias-voltage and the gate-voltage. Parameters are \revisiontwo{$k_\tx{B}T=U/200$, $\Delta=0.7U$, and $\Gamma_\tx{S}=\Gamma_\tx{D}\equiv\Gamma=10^{-5}U/\hbar$}. (b) Waiting time distributions corresponding to the points in the conductance map.}\label{fig:fig3}
\end{figure*}

\textit{Transport setup.}--- \revision{Having understood the importance of higher-order tunneling processes, we go on to consider a transport setup, where a bias-voltage between two electrodes drives a current through the quantum dot. We consider the waiting time between electrons that tunnel into the drain electrode and evaluate the waiting time distribution up to second order in the equal couplings to the source and the drain electrodes, $\Gamma_\tx{S}=\Gamma_\tx{D}$~\cite{SM}.} Figure~\ref{fig:fig3}(a) shows a conductance map  as a function of the bias-voltage $V$ and the gate voltage, which controls the energy level $\varepsilon$ of the quantum dot. In the low-bias regime \revisionthree{(green circle)}, Coulomb blockade suppresses the current in the quantum dot, and elastic cotunneling is the dominant transport mechanism. With an increasing voltage, additional transport processes are activated; first inelastic cotunneling \revisionthree{(orange square)}, then sequential tunneling \revisionthree{(blue triangle)}, and eventually all excitations are inside the bias window such that the Coulomb interactions effectively become irrelevant \revisionthree{(red star)}.

Figure~\ref{fig:fig3}(b) shows waiting time distributions corresponding to the four points in the conductance map. \revision{At low voltages, the waiting time distribution is suppressed at short times, showing that elastic cotunneling processes rarely occur simultaneously. In this case, the quantum dot mainly makes transitions between being occupied by a spin-down electron and being doubly occupied.  As  the  voltage  is  increased,  the  suppression at short times is lifted as inelastic cotunneling  gets activated, and the quantum dot can now make transitions from the doubly occupied state to being occupied by a spin-up electron, which quickly leaves via the drain electron before the quantum dot is refilled again. These fast events give rise to the peak at short waiting times, while the long waiting times occur when the quantum dot is occupied by a spin-down electron.  As the voltage is further increased, sequential  tunneling  becomes  the  main  transport  mechanism, which can be captured by a classical description as we have checked. Thus, we find that the waiting time distributions can be explained with a classical theory for large bias voltages, while our quantum theory is needed to capture the temporal fluctuations inside the Coulomb blockade region for small voltages and temperatures.
	
\revisiontwo{\textit{Experimental perspectives.}--- Finally, we comment on realistic experimental parameters and the perspectives for measuring the waiting time distributions found here. Taking an interacting strength of about $U\simeq 50\, \mu \text{eV}$, the tunneling rate in Fig.~\ref{fig:fig1} would be about $\Gamma\simeq 10\,\text{kHz}$ and the temperature  around $T\simeq 30\,\text{mK}$. These parameters are compatible with recent measurements of waiting time distributions \cite{Brange2021} and the real-time detection of single-electron tunneling involving virtual processes~\cite{Gustavsson_2008}. Moreover, state-of-the-art setups can detect single electrons on a microsecond timescale~\cite{Bauerle_2018}, corresponding to tunneling rates on the order of $\Gamma\simeq 1\, \text{MHz}$ as in Fig.~\ref{fig:fig3} for $U\simeq 50\, \mu \text{eV}$ and $T\simeq 10\,\text{mK}$. In addition, we expect  further improvements of charge detectors with bandwidths that are one or two orders of magnitude larger.}

\textit{Conclusions.}--- We have investigated the temporal fluctuations of charge tunneling in a quantum dot and developed a real-time diagrammatic theory of electron waiting time distributions that enables a systematic description of interaction effects and higher-order tunneling processes.} Our theoretical framework bridges the gap between non-interacting theories of electron waiting times, valid for phase-coherent transport in mesoscopic conductors, and master-equation descriptions, which apply to sequential tunneling in nano-structures with strong interactions. \revision{We have found that our quantum mechanical theory captures higher-order tunneling processes at low temperatures, which are not included in a classical description, and which dramatically affect the distribution of waiting times, for example by allowing fast tunneling processes inside the Coulomb blockade region of a quantum dot.} Our work paves the way for future investigations of waiting time distributions in systems with strong correlations, for example close to a Kondo resonance \cite{Komnik2005,Schmidt2007} or in a Luttinger liquid \cite{Komnik2006,Gutman2010}.

\textit{Acknowledgements.}--- We thank A.~Braggio for useful discussions. This work was supported by the German Research Foundation (DFG) within the Collaborative Research Centre (SFB) 1242 ``Non-Equilibrium Dynamics of Condensed Matter in the Time Domain" (Project No.~278162697) and the Ministry of Innovation NRW via the ``Programm zur F\"orderung der R\"uckkehr des Hochqualifizierten Forschungsnachwuchses aus dem Ausland". The work of CF was supported by Academy of Finland through the Finnish Centre of Excellence in Quantum
Technology (project nos.~312057 and 312299) and
project no.~308515. PS acknowledges support from the German National Academy of Sciences Leopoldina (Grant No.~LPDS 2019-10).

\end{document}

\title{Supplemental Material for\\
Electron Waiting Times in a Strongly Interacting Quantum Dot:\\ Interaction Effects and Higher-Order Tunneling Processes}
 
\author{Philipp Stegmann}\email{philipp.stegmann@uni-due.de}
\affiliation{Theoretische Physik, Universit\"at Duisburg-Essen and CENIDE, 47048 Duisburg, Germany}

\author{Bj\"orn Sothmann}
\affiliation{Theoretische Physik, Universit\"at Duisburg-Essen and CENIDE, 47048 Duisburg, Germany}

\author{J\"urgen K\"onig}
\affiliation{Theoretische Physik, Universit\"at Duisburg-Essen and CENIDE, 47048 Duisburg, Germany}

\author{Christian Flindt}
\affiliation{Department of Applied Physics, Aalto University, 00076 Aalto, Finland}

%\date{\today}

\maketitle

\onecolumngrid

\renewcommand{\theequation}{S\arabic{equation}}%
\renewcommand{\thefigure}{S\arabic{figure}}%

\section{Generalized master equation and real-time diagrams}\label{sec:mastereq}

We consider a generic transport setup determined by the full Hamiltonian $\mathcal{\hat H}=\mathcal{\hat H}_\tx{res}+\mathcal{\hat H}_\tx{qs}+\mathcal{\hat H}_\tx{tun}$. The leads are reservoirs of noninteracting electrons $\mathcal{\hat H}_\tx{res}=\sum_{\vec k \sigma r} \varepsilon_{\vec k \sigma r} \hat a^{\dagger}_{r \vec k \sigma } \hat a_{r \vec k \sigma}$, where $a^{\dagger}_{r \vec k \sigma}$ creates an electron with wave vector~$\vec k$, spin~$\sigma$, and energy~$\varepsilon_{r \vec k \sigma}$ in the reservoir $r$. The quantum system of interacting electrons is described by $\mathcal{\hat H}_\tx{qs}=\sum_\chi E_{\chi} \ket{\chi} \bra{\chi}$ with many-body eigenstates $\ket{\chi}$ at eigenenergies $E_{\chi}$. The quantum system and the reservoir are coupled via a tunneling Hamiltonian $\mathcal{\hat H}_{\tx{tun}}=\sum_{r \vec k  \sigma m \sigma'} g_{r \vec k \sigma  m \sigma' }\, \hat a_{r \vec k \sigma}^{\dagger} \hat d_{m \sigma'} + \text{H.c.}$, where $g_{r \vec k \sigma m \sigma' }$ is the tunnel matrix element and $\hat d_{m \sigma'}$ annihilates an electron of the quantum system in orbital $m$ with spin $\sigma'$.

We are interested in the time evolution of the reduced density matrix of the quantum system
\begin{equation}
\hat \rho(t)=\tx{Tr}_\tx{res}[\hat \rho_\tx{full}(t)]=\sum_{\chi_1,\chi_2}\rho_{\chi_2}^{\chi_1}(t) \ket{\chi_1}\bra{\chi_2}\, ,
\end{equation}
which is obtained from the density matrix of the full system $\hat \rho_\tx{full}$ by tracing out the reservoir degrees of freedom. We assume that at $t_0<t$ the full system is in a separable state $\hat \rho_\tx{full}(t_0)= \hat \rho (t_0) \otimes \hat \rho_\tx{res}(t_0)$ and the reduced density matrix of the leads $\hat \rho_\tx{res}(t_0)$ is in thermodynamic equilibrium.

The time evolution of the reduced density matrix can formally be written as
\begin{equation}\label{eq:rho}
\rho_{\chi_2}^{\chi_1}(t) = \Big\langle (\ket{\chi_2} \bra{\chi_1})(t) \Big\rangle =\sum_{\chi_1', \chi'_2} \rho_{\chi'_2}^{\chi'_1} (t_0) \Pi^{\chi '_1 \chi_1}_{\chi '_2 \chi_2}(t_0,t)
\end{equation}
with the \textit{full propagator} in the interaction picture
\begin{equation}
\Pi^{\chi '_1 \chi_1}_{\chi '_2 \chi_2}(t_0,t)=  \tx{Tr}_\tx{res} \Bigl[\hat \rho_\tx{res}(t_0) \bra{\chi_2'} \hat U_\tx{I} (t_0,t) (\ket{\chi_2} \bra{\chi_1}) (t)_\tx{I}  \hat U_\tx{I} (t,t_0)\ket{\chi_1'} \Bigr]
=\tx{Tr}_\tx{res} \Bigl[ \hat \rho_\tx{res}(t_0) \bra{\chi_2'} T_\tx{K} \hat U_\tx{K} (\ket{\chi_2} \bra{\chi_1}) (t)_\tx{I} \ket{\chi_1'} \Bigr]\, .
\end{equation}
The full propagator describes the time evolution from the matrix element $\rho_{\chi'_2}^{\chi'_1} (t_0)$ to $\rho^{\chi_1}_{\chi_2}(t)$. Formulated on the \textit{Keldysh contour} K, it gives the time evolution from state $\chi_1'$ at time $t_0$ to $\chi_1$ at $t$ and then from $\chi_2$ at $t$ backwards to $\chi_2'$ at $t_0$.
The Keldysh time-ordering operator $T_\tx{K} $ ensures the correct order of the operators along the Keldysh contour. Operators later in the Keldysh time appear further left. The Keldysh time-evolution operator $U_\tx{K}$ can be expanded in powers of the tunneling Hamiltonian
\begin{equation}\label{eq:Ukexp}
T_\tx{K} \hat U_\tx{K} := T_\tx{K} \exp \left( -\frac{i}{\hbar}\int\limits_\tx{K}  dt' \mathcal{\hat H}_\tx{tun}(t')_\tx{I} \right)= \sum_{n=0}^{\infty} \underset{t_1\ge t_2\ge \cdots \ge t_{2n}}{\int\limits_\tx{K}  dt_1\int\limits_\tx{K}  dt_2 \cdots \int\limits_\tx{K}  dt_{2n}}   \left(-\frac{i}{\hbar} \right)^{2n} T_\tx{K}  \mathcal{\hat H}_{\tx{tun}}(t_1)_\tx{I}  \mathcal{\hat H}_{\tx{tun}}(t_2)_\tx{I}  \cdots \mathcal{\hat H}_{\tx{tun}}(t_{2n})_\tx{I}\, ,
\end{equation}
with $t_1\ge t_2\ge \cdots \ge t_{2n}$ in the sense of the Keldysh contour. Since $\mathcal{\hat H}_\tx{res}$ is bilinear in the leads' electron operators, we can apply Wick's theorem~\cite{mahan_many_2000} to calculate the trace over the reservoir degrees of freedom. Operators $\hat a_{\vec k \sigma r}^{\dagger}$ and $\hat a_{\vec k \sigma r}$ are \textit{contracted} in time-ordered pairs, which is also the reason why \eq{Ukexp} includes only even orders of the tunneling Hamiltonian. In contrast, $\mathcal{\hat H}_\tx{qs}$ describes interacting electrons and is, therefore, treated explicitly. We insert identities $\hat 1 = \sum_\chi \ket{\chi}\bra{\chi}$ between the tunneling Hamiltonians and perform the remaining calculations in the Schr\"odinger picture.
In conclusion, each time-ordered product of order $n$
\begin{equation}\nonumber
\tx{Tr}_\tx{res}\left[\hat \rho_\tx{res}(t_0) \bra{\chi_2'} \mathcal{\hat H}_{\tx{tun}}(t_1)_\tx{I}  \mathcal{\hat H}_{\tx{tun}}(t_2)_\tx{I} \cdots (\ket{\chi_2} \bra{\chi_1}) (t)_\tx{I} \cdots \mathcal{\hat H}_{\tx{tun}}(t_{2n})_\tx{I}  \ket{\chi_1'} \right]
\end{equation}
is calculated by summing over all possible \textit{trajectories}, i.e., all combinations of contractions and projectors $\ket{\chi}\bra{\chi}$ leading to a nonvanishing value.
\begin{figure}[t]
	\includegraphics[scale=0.88]{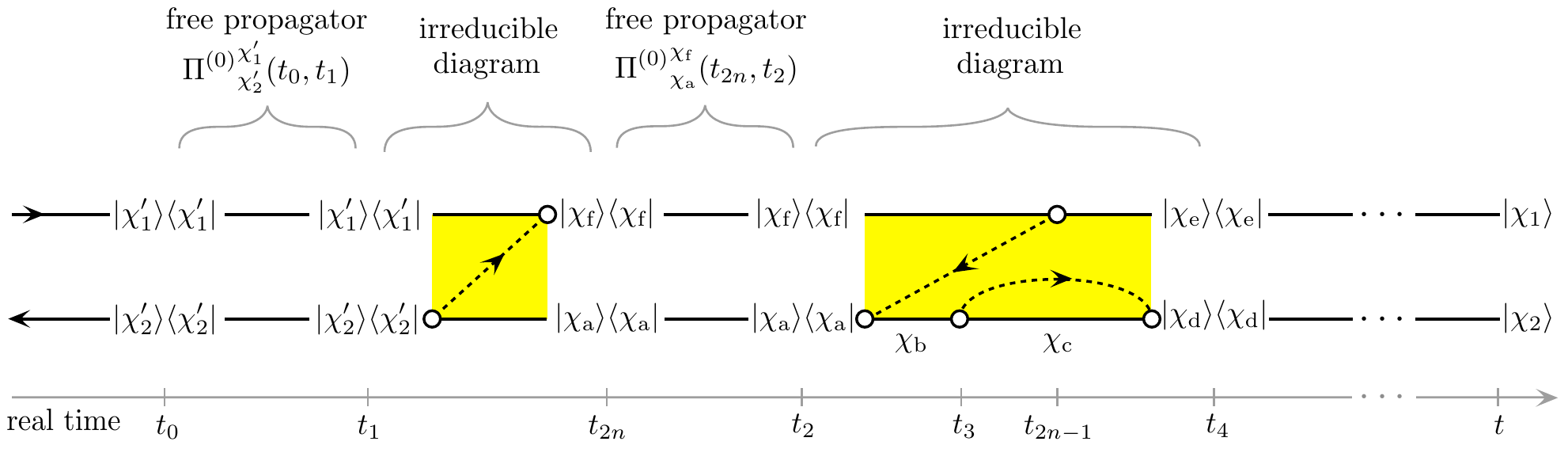}
	\caption{Real-time dynamics on the Keldysh contour from $t_0$ to $t$ broken down in free propagators and irreducible diagrams. The illustrated trajectory contributes to $\tx{Tr}_\tx{res}\left[\hat \rho_\tx{res}(t_0) \bra{\chi_2'} \mathcal{\hat H}_{\tx{tun}}(t_1)_\tx{I}  \mathcal{\hat H}_{\tx{tun}}(t_2)_\tx{I} \cdots (\ket{\chi_2} \bra{\chi_1}) (t)_\tx{I} \cdots \mathcal{\hat H}_{\tx{tun}}(t_{2n})_\tx{I}  \ket{\chi_1'} \right]$.
	}\label{fig:fig1}
\end{figure}

Each trajectory can be visualized as illustrated in \fig{fig1}. Black lines represent free propagating parts of the Keldysh contour, i.e., $\bra{\chi}U_0(t',t)\ket{\chi}$ with $\hat U_0(t',t)= \exp[-i (\hat H_\tx{res}+\hat H_\tx{qs})(t'-t)/\hbar]$. Contractions are indicated as dashed \textit{tunneling lines}. The open circles on the contour are called \textit{tunneling vertices}. Vertices with a leaving (entering) tunneling line give $\bra{\chi'} (-i/\hbar)  g_{r \vec k \sigma  m \sigma' }\, \hat a_{r \vec k \sigma}^{\dagger} \hat d_{m \sigma'} \ket{\chi}$ $\left(\bra{\chi'} (-i/\hbar)  g^*_{r \vec k \sigma  m \sigma' }\,  \hat d^\dagger_{m \sigma'} \hat a_{r \vec k \sigma} \ket{\chi}\right)$, where $\chi$ are the entering and $\chi'$ the leaving states. Thus, the emission (absorption) of an electron by the quantum system is indicated by tunneling lines which leave (enter) the vertices. 

The time evolution from $t_0$ and $t$ is separated in \textit{free propagators}
\begin{equation}
{\Pi^{(0)}}^{\chi_1}_{\chi_2}(t',t)=\bra{\chi_2}U_0(t',t)\ket{\chi_2}\bra{\chi_1}U_0(t,t')\ket{\chi_1}=e^{-\frac{i}{\hbar}(E_{\chi_1}-E_{\chi_2})(t-t')}
\end{equation}
and \textit{irreducible diagrams} which describe tunneling events between the leads and the quantum system. Irreducible means here that one can not draw a vertical line through these diagrams without cutting a tunneling line.

From the form of the trajectories depicted in \fig{fig1}, it is immediately clear that we can reformulate the full propagator recursively in the sense of a Dyson equation
\begin{equation}\label{eq:fullpropagatortime}
\Pi^{\chi '_1 \chi_1}_{\chi '_2 \chi_2}(t_0,t) = \hspace{0.15cm} {\Pi^{(0)}}^{\chi_1}_{\chi_2}(t_0,t) \delta_{\chi '_1, \chi_1}  \delta_{\chi '_2, \chi_2} 
+\sum_{\chi ''_1 ,\chi_2 ''}\,\int\limits^{t}_{t_0}d{t_2}\int\limits^{t_2}_{t_0}d{t_1}\Pi^{\chi '_1 \chi_1''}_{\chi '_2 \chi_2''}(t_0,t_1)\mathbb{W}^{\chi_1'' \chi_1}_{\chi_2'' \chi_2}(t_1,t_2){\Pi^{(0)}}^{\chi_1}_{\chi_2}(t_2,t).
\end{equation}
The integral kernel $\mathbb{W}^{\chi_1' \chi_1}_{\chi_2' \chi_2}(t',t)$ is the sum of all possible irreducible diagrams connecting the matrix element $\rho^{\chi_1'}_{\chi_2'}(t')$ with $\rho^{\chi_1}_{\chi_2}(t)$. Hence, the leftmost vertex of each irreducible diagram is at $t'$ and the rightmost vertex is at $t$.

Inserting \eq{fullpropagatortime} into \eq{rho} and differentiating with respect to $t$ yields the \textit{generalized master equation}
\begin{equation}\label{eq:master}
\frac{d}{d{t}}\rho^{\chi_1}_{\chi_2}(t)=-\frac{i}{\hbar}(E_{\chi_1}-E_{\chi_2})\rho^{\chi_1}_{\chi_2}(t)+\sum_{\chi_1',\chi_2'}\, \int\limits^t_{t_0}d{t'}\rho^{\chi_1'}_{\chi_2'}(t')\mathbb{W}^{\chi_1' \chi_1}_{\chi_2' \chi_2}(t',t).
\end{equation}
The first term on the right-hand side describes the time evolution in the absence of tunneling events. It arises from the free propagator in \eq{fullpropagatortime}. The second term determines the dissipative coupling to the leads. This term forces the quantum system to approach a stationary state. The kernel in \eq{master} is written down as function of both $t'$ and $t$. However, since the system is not explicit time dependent, the kernel depends only on the time difference $t-t'$ and we can write $\mathbb{W}^{\chi_1' \chi_1}_{\chi_2' \chi_2}(t',t)=\mathbb{W}^{\chi_1' \chi_1}_{\chi_2' \chi_2}(t-t')$.

At last, we introduce a detector which does not change the dynamics of the system but counts the number of electrons $N$ that have been transferred through the quantum system in time interval $[0,t]$. 
We replace the master equation for $\rho^{\chi_1}_{\chi_2}(t)$ by an $N$-resolved version for ${\rho_N}^{\chi_1}_{\chi_2}(t)$
\begin{equation}\label{eq:mastercount}
\frac{d}{d{t}}{\rho_N}^{\chi_1}_{\chi_2}(t)=-\frac{i}{\hbar}(E_{\chi_1}-E_{\chi_2}){\rho_N}^{\chi_1}_{\chi_2}(t)+\sum_{N',\chi_1',\chi_2'}\, \int\limits^t_{t_0}d{t'}{\rho_{N'}}^{\chi_1'}_{\chi_2'}(t')  {{{\widetilde{\mathbb{W}}}_{N-N'}} \hspace{0.0cm}^{\chi_1' \chi_1}_{\chi_2' \chi_2}}(t',t).
\end{equation}
The integral kernel ${{{\widetilde{\mathbb{W}}}_{N-N'}} \hspace{0.0cm}^{\chi_1' \chi_1}_{\chi_2' \chi_2}}(t',t)$ includes only those irreducible diagrams that change the detector counter by $N-N'$. The unresolved density matrix and kernel are recovered by summing over $N$, i.e., ${\rho}^{\chi_1}_{\chi_2}(t)=\sum_N {\rho_N}^{\chi_1}_{\chi_2}(t)$ and ${{{{\mathbb{W}}}} \hspace{0.0cm}^{\chi_1' \chi_1}_{\chi_2' \chi_2}}(t',t)=\sum_N{{{\widetilde{\mathbb{W}}}_{N-N'}} \hspace{0.0cm}^{\chi_1' \chi_1}_{\chi_2' \chi_2}}(t',t)$. Time translation invariance is broken at $t'=0$ due to the switching on of the detector. Therefore, the time dependence of the kernel simplifies to ${{{\widetilde{\mathbb{W}}}_{N-N'}} \hspace{0.0cm}^{\chi_1' \chi_1}_{\chi_2' \chi_2}}(t',t)= {{{{\mathbb{W}}}_{N-N'}} \hspace{0.0cm}^{\chi_1' \chi_1}_{\chi_2' \chi_2}}(t-t')$ only for $t'\ge 0$. Moreover, ${\rho_{N'}}^{\chi_1}_{\chi_2}(t')=\delta_{N'\!,0}\,  {\rho}^{\chi_1}_{\chi_2}(t')$ for $t'<0$. Thus, \eq{mastercount} takes the form
\begin{equation}\label{eq:masternfin}
\frac{d}{d{t}}{\rho_N}^{\chi_1}_{\chi_2}(t)=-\frac{i}{\hbar}(E_{\chi_1}-E_{\chi_2}){\rho_N}^{\chi_1}_{\chi_2}(t)+\sum_{N',\chi_1',\chi_2'}\, \int\limits^t_{0}d{t'}{\rho_{N'}}^{\chi_1'}_{\chi_2'}(t')  {{{{\mathbb{W}}}_{N-N'}} \hspace{0.0cm}^{\chi_1' \chi_1}_{\chi_2' \chi_2}}(t-t')+\sum_{\chi_1',\chi_2'}\, \int\limits^0_{t_0}d{t'}{\rho}^{\chi_1'}_{\chi_2'}(t')  {{{\widetilde{\mathbb{W}}}_{N}} \hspace{0.0cm}^{\chi_1' \chi_1}_{\chi_2' \chi_2}}(t',t).
\end{equation}
The last term is called \textit{inhomogeneity} ${\gamma_N}^{\chi_1}_{\chi_2}(t)$. The kernel ${{{\widetilde{\mathbb{W}}}_{N}} \hspace{0.0cm}^{\chi_1' \chi_1}_{\chi_2' \chi_2}}(t',t)$ is nonvanishing for $t-t'$ being smaller than the memory time of the kernel. If we assume that the system has reached its stationary state earlier, we can set $t_0 \to -\infty$ and ${\rho}^{\chi_1'}_{\chi_2'}(t') \to {\rho_\text{stat}}^{\chi_1'}_{\chi_2'}$. 
In conclusion, we have recovered Eq.~(2) of the main text. There, we concentrated on systems where the density matrix is diagonal and the first term in \eq{masternfin} vanishes.

\section{Diagrammatic rules and counting fields}

\begin{figure}[t]
	\includegraphics[scale=0.88]{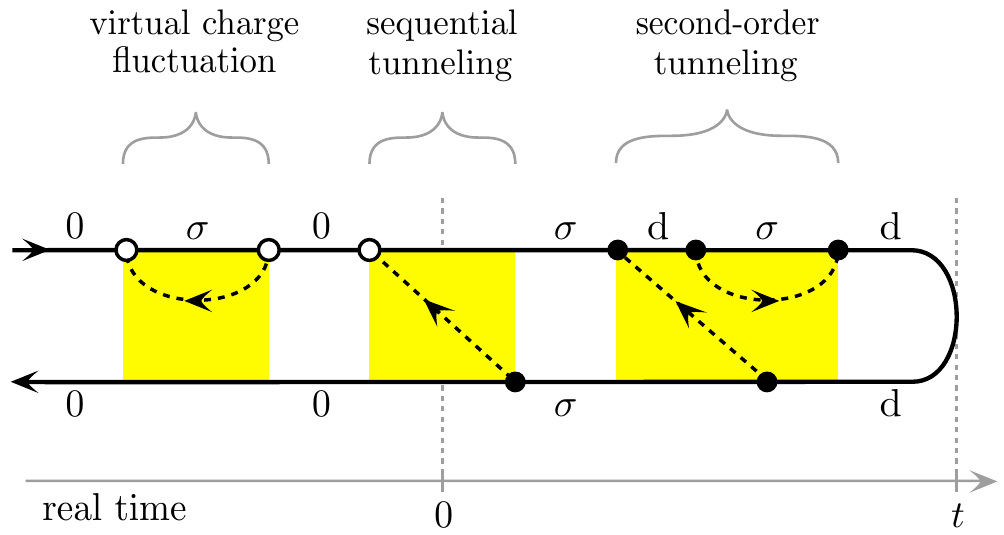}
	\caption{Real-time dynamics on the Keldysh contour for the Hamiltonian studied in the main text. Counted tunneling vertices after $t=0$ are shown as filled black circles, while earlier ones are indicated by open circles.
	}\label{fig:count}
\end{figure}

In this section, we present rules how to calculate the kernel elements ${\mathbb{W}}_{\! s}^{(n)} \hspace{0.0cm}^{\chi_1' \chi_1}_{\chi_2' \chi_2}(z)$ for the Hamiltonian studied in the main text. The superscript $(n)$ indicates the order in the tunnel-coupling strength $\Gamma_r=2\pi | g_r |^2 \nu_{r}/\hbar$, where $\nu_r$ is the density of states in lead $r$. The Fermi function of lead $r$ is $f^+_r(\omega)=1/[e^{(\omega-\mu_r)/k_\text{B}T}+1]$ and $f^-_r(\omega)=1-f^+_r(\omega)$.
Since counting begins at $t=0$, only tunneling vertices later in real time, indicated by filled black circles in \fig{count}, contribute to the detector counter. We follow the counting procedure introduced in Ref.~\onlinecite{levitov_counting_2004} and apply the replacement $g_r \to g_r e^{i X_r/2}$ to filled tunneling vertices on the upper and $g_r \to g_r e^{-i X_r/2}$ to the ones on the lower Keldysh contour. Afterwards, we change from counting fields to counting factors via the replacement $i X_r \to \ln s_r$. The resulting counting procedure is summarized in the 4th rule.
More details of the counting procedure are given at the beginning of \Sec{kernelfig3}.
The rules are:

\begin{enumerate}
\item
Draw all topologically different irreducible diagrams with $n$ tunneling lines connecting tunneling vertices on either the same or opposite contours. Assign the states $\chi_1' $, $ \chi_1$, $\chi_2'$, and $\chi_2$ to the four corners of the diagram and the respective states to the ends of each free propagating part. Furthermore, each free propagating part gets its corresponding energy $E_\chi$ as well as each tunneling line its energy $\omega$.

\item
For each time interval on the real axis confined by two neighboring vertices (independent if they are on the same or opposite contours), write down a resolvent $1/(\Delta E +i z)$, where $\Delta E$ is the energy of the left-going free propagating parts and tunneling lines minus the energy of the right-going ones. 

\item
Add a factor $\frac{\hbar}{2\pi} f^{-}_r(\omega)$ for each tunneling line running forward and $ \frac{\hbar}{2\pi} f^{+}_r(\omega)$ for each tunneling line running backwards in the Keldysh time.

\item
For each pair of vertices connected by a tunneling line, multiply with
\[ 
s_r^{m/2} \Gamma_r \bra{\chi_4'} \hat d_{\uparrow} \ket{\chi_4}\bra{\chi_3'} \hat d^{\dagger}_{\uparrow} \ket{\chi_3}+s_r^{m/2}  \Gamma_r \bra{\chi_4'}  \hat d_{\downarrow} \ket{\chi_4}\bra{\chi_3'} \hat d^{\dagger}_{\downarrow} \ket{\chi_3},
\]
where $\chi_3$ and $\chi_4$ ($\chi_3'$ and $\chi_4'$) are the states that enter (leave) the vertex with respect to the Keldysh time. Here, $m$ is the number of filled vertices with beginning tunneling line on the upper or ending tunneling line on the lower contour minus the number of filled vertices with ending tunneling line on the lower or beginning tunneling line on the upper contour.

\item
Assign the factor $-\frac{i}{\hbar}(-1)^{b+c}$ where $b$ is the number of vertices on the lower contour and $c$ the number of crossings of tunneling lines.

\item
Integrate over all energies $\omega$ and sum over the leads $r$.

\item
Sum up the contributions of all diagrams.

\end{enumerate}

\section{Diagrammatic Kernel}\label{sec:kernelfig3}
We present the kernel elements obtained from the diagrammatic rules and used to produce Fig.~1, 2, 3(b), and \ref{fig:fig3}. We sum over $r$ and $r'$ in the expression given below. Afterwards, we set $\Gamma_\tS=0$, $s_\tD^{-1}=1$, and $s_\tD=s$ in case of Fig.~1, 2, and \ref{fig:fig3}. 
For the waiting time distribution in Fig.~3(b), we set $s_\tD^{-1}=s_\tS^{-1}=s_\tS=1$ and  $s_\tD=s$.

The nonvanishing kernel elements in first order in the tunnel-coupling strength read
\begin{align}
{\mathbb{W}}_{\! s_\tS s_\tD}^{(1)}\hspace{0.0cm}^{0 \sigma}_{0 \sigma}(z=0) &= \sum_{r=\tS,\tD} \frac{1}{s_r} \Gamma_r f^+_r(\varepsilon_\sigma)\, , \label{eq:sq1}\\
{\mathbb{W}}_{\! s_\tS s_\tD}^{(1)}\hspace{0.0cm}^{\sigma 0}_{\sigma 0}(z=0) &= \sum_{r=\tS,\tD} s_r \Gamma_r f^-_r(\varepsilon_\sigma)\, ,\label{eq:sq2}\\
{\mathbb{W}}_{\! s_\tS s_\tD}^{(1)}\hspace{0.0cm}^{\sigma \tx{d}}_{\sigma \tx{d}}(z=0) &= \sum_{r=\tS,\tD} \frac{1}{s_r} \Gamma_r  f^+_r(\varepsilon_{\bar \sigma}+U)\, , \label{eq:sq3}\\
{\mathbb{W}}_{\! s_\tS s_\tD}^{(1)}\hspace{0.0cm}^{\tx{d} \sigma}_{\tx{d} \sigma}(z=0) &= \sum_{r=\tS,\tD} s_r \Gamma_r  f^-_r(\varepsilon_{\bar \sigma}+U)\, ,\label{eq:sq4}\\
{\mathbb{W}}_{\! s_\tS s_\tD}^{(1)}\hspace{0.0cm}^{\chi \chi}_{\chi \chi}(z=0) &= -\sum_{\chi' \neq \chi} {\mathbb{W}}_{\! 11}^{(1)}\hspace{0.0cm}^{\chi \chi'}_{\chi \chi'}(z=0)\, , \label{eq:sq5}
\end{align}
with $\varepsilon_\uparrow=\varepsilon+\frac{\Delta}{2}$ and $\varepsilon_\downarrow=\varepsilon-\frac{\Delta}{2}$.
The elements given in Eqs.~(\ref{eq:sq1}) - (\ref{eq:sq4}) account for sequential tunneling processes from state $\chi'$ to state $\chi$ where one electron is exchanged between the leads and the dot. The kernel elements given in \eq{sq5} are determined by virtual charge fluctuations. They ensure the conservation of probability.

The kernel elements which are relevant in the elastic and inelastic cotunneling regime are
\begin{align}
{\mathbb{W}}_{\! s_\tS s_\tD}^{(2)}\hspace{0.0cm}^{\downarrow 0}_{\downarrow 0}(z=0) &=  \sum_{r'r} s_{r'} {\Gamma^{\text{sq}}_\tx{ren\,}}_{\downarrow 0}^{r'r}\, , \\
{\mathbb{W}}_{\! s_\tS s_\tD}^{(2)}\hspace{0.0cm}^{\downarrow \uparrow}_{\downarrow \uparrow}(z=0) &=  \sum_{r'r} \frac{s_{r'}}{s_{r}} \left( {\Gamma^{\text{cot}}}_{\downarrow \uparrow}^{r'r} + {\Gamma^{\text{cot}}_\tx{ren}}_{\downarrow \uparrow}^{r'r}  \right)\ ,\\
{\mathbb{W}}_{\! s_\tS s_\tD}^{(2)}\hspace{0.0cm}^{\downarrow \downarrow}_{\downarrow \downarrow}(z=0) &=  \sum_{r'r} \left(\frac{s_{r'}}{s_{r}}-1\right) \left( {\Gamma^{\text{cot}}}_{\downarrow \downarrow}^{r'r} + {\Gamma^{\text{cot}}_\tx{ren}}_{\downarrow \downarrow}^{r'r}  \right) - \sum_{\chi \neq \downarrow} \, {\mathbb{W}}_{\! 11}^{(2)}\hspace{0.0cm}^{\downarrow \chi}_{\downarrow \chi}(z=0)\, , \\
{\mathbb{W}}_{\! s_\tS s_\tD}^{(2)}\hspace{0.0cm}^{\downarrow \tx{d}}_{\downarrow \tx{d}}(z=0) &=  \sum_{r'r} \frac{1}{s_{r'}}  {\Gamma^{\text{sq}}_\tx{ren\,}}_{\downarrow \tx{d}}^{r'r}\, .
\end{align}
The individual terms involved take the form
\begin{align}
{\Gamma^{\text{cot}}}_{\downarrow \downarrow}^{r'r}&= \partial_x \gamma(\mu_{r'},\mu_r,U,0) \, ,\label{eq:elcot}\\
{\Gamma^{\text{cot}}}_{\downarrow \uparrow}^{r'r}&=\partial_x \gamma(\mu_{r'},\mu_r-\Delta,U-\Delta,0) +\frac{2}{U}\gamma(\mu_{r'},\mu_r-\Delta,U-\Delta,0)\, ,\label{eq:incot}\\
{\Gamma^{\text{sq}}_\tx{ren\,\,}}_{\downarrow 0}^{r'r}&=-\hbar \Gamma_{r'} \Gamma_{r} \partial \phi_{r'}(\varepsilon_\downarrow) \, \left[f^+_r(\varepsilon_\uparrow)+ f^+_r(\varepsilon_\downarrow)\right] \, , \label{eq:rendown0} \\
{\Gamma^{\text{cot}}_\tx{ren\,}}_{\downarrow \uparrow}^{r'r}&=\hbar \Gamma_{r'} \Gamma_{r} \partial \phi_{r'}(\varepsilon_\downarrow) \, f^+_r(\varepsilon_\uparrow) -  \hbar \Gamma_{r'} \Gamma_{r} \partial \phi_{r'}(\varepsilon_\uparrow+U) f^-_r(\varepsilon_{\downarrow}+U)\, ,  \\
{\Gamma^{\text{cot}}_\tx{ren\,}}_{\downarrow \downarrow}^{r'r}&=\hbar \Gamma_{r'} \Gamma_{r} \partial \phi_{r'}(\varepsilon_\downarrow) \, f^+_r(\varepsilon_\downarrow) -  \hbar \Gamma_{r'} \Gamma_{r} \partial \phi_{r'}(\varepsilon_\uparrow+U) f^-_r(\varepsilon_{\uparrow}+U)\,, \\
{\Gamma^{\text{sq}}_\tx{ren\,}}_{\downarrow \tx{d}}^{r'r}&=\hbar \Gamma_{r'} \Gamma_{r} \partial \phi_{r'}(\varepsilon_\uparrow+U) \, \left[f^-_r(\varepsilon_{\uparrow}+U) +f^-_r(\varepsilon_{\downarrow}+U) \right] \, , \label{eq:gamren}
\end{align}
with $\phi_r(x)=\frac{1}{2}\text{Re} \Psi(\frac{1}{2}+i\frac{x-\mu_r}{2\pi k_\text{B}T})$, where $\Psi(x)$ is the digamma function. The elastic and inelastic cotunneling rates ${\Gamma^{\text{cot}}}_{\downarrow \downarrow}^{r'r}$ and ${\Gamma^{\text{cot}}}_{\downarrow \uparrow}^{r'r}$ are determined by
\begin{equation}
\gamma(\mu_{r'},\mu_r,U,x)=\hbar\Gamma_{r'}\Gamma_r f^\text{B}_r(\mu_{r'})\left[\phi_{r'}\left(\varepsilon_\downarrow+x\right)-\phi_{r'}\left(\varepsilon_\uparrow+U-x\right)-\phi_r\left(\varepsilon_\downarrow+x\right)+\phi_r\left(\varepsilon_\uparrow+U-x\right)\right]\, ,
\end{equation}
with the Bose function $f^\text{B}_r(x)=1/[e^{(x-\mu_{r})/k_\text{B}T}-1]$. The remaining terms in Eqs.~(\ref{eq:rendown0}) - (\ref{eq:gamren}) are second-order corrections due to a renormalized tunnel-coupling strength. 

\section{Inhomogeneity}\label{sec:inhom}
The kernel $\widetilde{\mathbb{W}}_{\! s}(t,t')$ is build up by probabilities that certain numbers of tunneling vertices occur in the time span~$[t',0]$ and $[0,t]$. 
The calculation simplifies in leading order in the tunnel-coupling strength. Then, one vertex occurs at $t'<0$ (not counted) and a second one at $t>0$ (counted). The kernel $\widetilde{\mathbb{W}}_{\!s}^{(1)}(t,t')= \widetilde {\mathbb{W}}^{(1)}_{\!s}(t-t')$ depends only on the time difference between those two vertices. The inhomogeneity takes the form
\begin{equation}\label{eq:inhle}
	\hat \gamma^{(1)}_s(t) = \left[  \int\limits^{t}_{-\infty}dt'\, \widetilde{\mathbb{W}}^{(1)}_{\! s}(t-t') -   \int\limits^{t}_{0}dt' \,\widetilde{\mathbb{W}}^{(1)}_{\! s}(t-t') \right] \hat \rho^{(0)}_\text{stat}\, .
\end{equation}
Here, we have assumed that $\hat \rho_{N}(t\le 0)=\delta_{N,0}\, \hat \rho_\text{stat}$ and, moreover, extended the kernel $\widetilde {\mathbb{W}}^{(1)}_{\!s}(t-t')$ to the domain $0<t'<t$. The Laplace transform is
\begin{equation}\label{eq:gam1}
	\hat \gamma^{(1)}_s(z) =\frac{1}{z} \left[ \widetilde{\mathbb{W}}^{(1)}_{\! s}(0)-  \widetilde{\mathbb{W}}^{(1)}_{\! s}(z) \right]\hat \rho_\tx{stat}^{(0)}\, .
\end{equation}
We note that \eq{gam1} holds in arbitrary order if counting fields are not included ($s=1$)
\begin{equation}
	\hat \gamma_1(z) =\frac{1}{z} \left[ {\mathbb{W}}_{\! 1}(0)-  {\mathbb{W}}_{\! 1}(z) \right]\hat \rho_\tx{stat} =-\frac{1}{z}\, {\mathbb{W}}_{\! 1}(z) \,\hat \rho_\tx{stat} \, .
\end{equation}

\section{Waiting time distribution}

In this section, we derive the non-Markovian form of the waiting time distribution in detail. We start with Eq.~(3) of the main text
\begin{equation}
	\hat \rho_s(z)=\frac{1}{z-\mathbb{W}_{\! s}(z)}[\hat \rho_\text{stat} +\hat \gamma_s (z)]\, .
\end{equation}
The prefactor is written in the form of a Neumann series~$[1-\mathbb{W}_{\! s}(z)]^{-1}=\sum_{n=0}^\infty \mathbb{W}_{\! s}^n(z)/z^{n+1}$ and $\mathbb{W}^n_{\! s}(z)$ as well as $\hat \gamma_{s}(z)$ are expressed by their Taylor series. We obtain
\begin{equation}
	\hat \rho_s(z)= \sum_{m,n,k=0}^{\infty} \frac{z^{m+k-n-1}}{m!\, k!} \Biggl\{ \partial_z^m \left[ \mathbb{W}_{\! s}^n(z) \right ] \, \partial_z^k \left[ \hat \rho_{\text{stat}} + \hat \gamma_s(z) \right]  \biggr\}_{\! z=0} \, ,
\end{equation}
where $\partial_z^k \hat \rho_{\text{stat}} = \delta_{k,0} \hat\rho_{\text{stat}}$.
The inverse Laplace transform yields $1/z^{n+1} \to t^{n}/n!$ for the negative powers and $z^n \to \partial^n_t \delta(t)$ for the positive powers in $z$. We assume finite $t$ and, therefore, neglect terms including a Dirac delta function~$\delta(t)$ or its derivatives. Thus, the solution of the master equation reads
\begin{equation}
	\hat \rho_s(t)= \sum_{m,k=0}^{\infty} \frac{1}{m!\, k!} \Biggl\{ \partial_z^m \left[  \mathbb{W}_{\! s}^{m+k}(z)e^{\mathbb{W}_{\! s}(z)t} \right ] \, \partial_z^k \left[ \hat \rho_{\text{stat}} + \hat \gamma_s(z) \right]  \biggr\}_{\! z=0} \, ,
\end{equation}
which can be simplified by means of the Leibniz rule to
\begin{equation}
	\hat \rho_s(t) =  \sum_{m=0}^{\infty} \frac{1}{m!}  \partial_z^m \!  \Biggl\{ \mathbb{W}_{\! s}^{m}(z)e^{\mathbb{W}_{\! s}(z)t}\, \left[\hat \rho_\text{stat} + \hat \gamma_{s}(z)\right]  \biggr\}_{\! z=0}\, .
\end{equation}
The waiting time distribution $w(\tau)= \braket{\tau} \partial_\tau^2 \Pi(\tau)$ can be derived from the mean waiting time $\braket{\tau}=-1/[\partial_\tau \Pi(0)]$ and the idle-time probability $\Pi(\tau)=\text{Tr}[\hat \rho_{s=0}(\tau)]$. We take into account that $\tx{Tr}[\mathbb{W}_{\! 1}(z)\bullet ]=0$ and ${\hat \gamma}_{1}(z)=-\mathbb{W}_{\! 1}(z) \hat \rho_\tx{stat}/z$, see Section~\ref{sec:inhom}. The waiting time distribution reads
\begin{equation}
	w(\tau)= \sum_{m=0}^\infty \frac{\braket{\tau} }{m!} \partial_z^m \text{Tr}\Biggl\{ \mathcal{J}(z) \,  \mathbb{W}_{\! 0}^m(z) e^{\mathbb{W}_{\! 0}(z)\tau} \left[\mathcal{J}(z) \hat \rho_\text{stat} + z \hat \gamma_1(z) - \mathbb{W}_{\!0}(z) \hat \gamma_0(z) \right]\Biggr\}_{z=0} \, ,
\end{equation}
with the jump operator $\mathcal{J}(z):=\mathbb{W}_{\!1}(z)-\mathbb{W}_{\! 0}(z)$. Finally, we use the relation $(m+1)\partial_z^m [f(z)]_{z=0}=\partial_z^{m+1} [f(z)z]_{z=0}$ and
\begin{equation}
	\widetilde{\mathcal{J}} (z) \hat \rho_\text{stat} := \mathcal{J}(z)+ \int\limits_{0}^\infty dt \,z e^{-z t} \int\limits_{-\infty}^0 dt' \, [\widetilde{\mathbb{W}}_{\! 1}(t,t')-\widetilde{\mathbb{W}}_{\! 0}(t,t')] \hat \rho_\text{stat} = \mathcal{J}(z) + z \left[\hat \gamma_1(z)- \hat \gamma_0(z) \right]
\end{equation}
to obtain Eq.~(4) of the main text
\begin{equation}\label{eq:wfull}
	w(\tau)= \sum_{m=0}^\infty \frac{\braket{\tau} }{m!} \partial_z^m \text{Tr}\left\{ \mathcal{J}(z) \,  \mathbb{W}_{\! 0}^m(z) e^{\mathbb{W}_{\! 0}(z)\tau} \widetilde{\mathcal{J}} (z)  \hat \rho_\text{stat} \right\}_{z=0} \, .
\end{equation}

\section{Perturbative expansions}\label{sec:wexpansion}

In this section, we expand the waiting time distribution in the tunnel-coupling strength $\Gamma$
\begin{equation}
	w(\tau)= \braket{\tau} \sum_{n=0}^\infty \partial_\tau^2 \Pi^{(n)}(\tau) = \braket{\tau} \sum_{n=0}^\infty \frac{\Gamma^n}{n!}\, \partial_\Gamma^n \left[\partial_\tau^2 \Pi(\tau) \right]_{\Gamma=0}\, ,
\end{equation}
with the superscript $(n)$ indicating the order of the tunnel-coupling strength. The product $\Gamma \tau$ is treated as $\Gamma$-independent variable in the following. In practice, the infinite sum is truncated at a certain order $n_\text{max}$. To preserve the normalization of the waiting time distribution, the mean waiting time is truncated accordingly. Its expansion is directly obtainable from the expansion of $\partial_\tau^2\Pi(\tau)$ and is one order lower in the coupling strength
\begin{equation}
	\braket{\tau}_{n_\text{max}-1} =  \left(\int\limits_{0}^\infty \! d\tau\,\sum_{n=0}^{n_\text{max}} \partial_\tau^2 \Pi^{(n)}(\tau) \right)^{-1}= -\left(\sum_{n=0}^{n_\text{max}-1} \partial_\tau \Pi^{(n)}(0) \right)^{-1}\, .
\end{equation}
We set $n_\text{max}=2$ to obtain the waiting time distribution in the lowest nonvanishing order. We recover from \eq{wfull} the well-known Markovian form~\cite{carmichael_photoelectron_1989,brandes_waiting_2008}
\begin{equation}\label{eq:wle}
	w_1^\text{M}(\tau)= \braket{\tau}_{1}\! \! \sum_{n=0}^{n_\text{max}=2} \partial_\tau^2 \Pi^{(n)}(\tau) = \braket{\tau}_{1} \frac{\Gamma^2}{2!}\, \partial_\Gamma^2 \left[\partial_\tau^2 \Pi(\tau) \right]_{\Gamma=0}  =\frac{\text{Tr}\left[ \mathcal{J}^{(1)} e^{\mathbb{W}^{(1)}_{\! 0} \tau}\mathcal{J}^{(1)} \hat \rho_\text{stat}^{(0)}\right]}{\text{Tr}\left[\mathcal{J}^{(1)} \hat \rho_\text{stat}^{(0)}\right]}\, .
\end{equation}
Here, only first-order processes are involved (sequential tunneling). The Laplace variable $z$ is not written down explicitly. The exponential function $e^{\mathbb{W}^{(1)}_{\! 0} \tau}$ includes only powers of $\Gamma \tau$ and is, therefore, considered to be independent of $\Gamma$. 
If we include the next-to-leading order ($n_\text{max}=3$), second-order Markovian as well as non-Markovian processes are taken into account. The waiting time distribution reads
\begin{equation}
	\begin{aligned}
		\frac{w_2(\tau)}{\braket{\tau}_2} =\,\, & \text{Tr}\bigl[ \mathcal{J}^{(1)} e^{\mathbb{W}_{\! 0}^{(1)}\tau} \mathcal{J}^{(1)} \hat \rho_\text{stat}^{(0)}\bigr] + 
		\frac{\Gamma^3}{3!}\partial_\Gamma^3 
		\text{Tr}\bigl[ ( \mathcal{J}^{(1)}\!+ \mathcal{J}^{(2)})e^{(\mathbb{W}^{(1)}_{\! 0}+\, \mathbb{W}^{(2)}_{\! 0}) \tau} ( \mathcal{J}^{(1)}\!+\mathcal{J}^{(2)}) ( \hat \rho_\text{stat}^{(0)}+\hat \rho_\text{stat}^{(1)})\bigr]_{\Gamma=0} \\
		& +  \frac{\Gamma^3}{3!}\partial_\Gamma^3\, \partial_z \text{Tr}\bigl[ \mathcal{J}^{(1)} \mathbb{W}^{(1)}_{\! 0} e^{\mathbb{W}^{(1)}_{\! 0}\tau} \widetilde{\mathcal{J}}^{(1)}\hat \rho_\text{stat}^{(0)}  \bigr]_{\Gamma=0}\, .
	\end{aligned}
\end{equation}
The first two terms are Markovian and the third term is a non-Markovian correction indicated by the $z$-derivative. We introduce the auxiliary jump operator $\mathbb{J}^{(2)}=\mathcal{L}^{(2)}_1-\mathcal{L}^{(2)}_0$ with $\mathcal{L}^{(2)}_s=\mathbb{W}^{(2)}_{\!s}+(\partial_z \mathbb{W}^{(1)}_{\!s})\mathbb{W}^{(1)}_{\!s}$ and obtain the form
\begin{equation}\label{eq:wleadnext}
	\begin{aligned}
		\frac{w_2(\tau)}{\braket{\tau}_2} =\,\, & \text{Tr}\bigl[ \mathcal{J}^{(1)} e^{\mathbb{W}_{\! 0}^{(1)}\tau} \mathcal{J}^{(1)}( \hat \rho_\text{stat}^{(0)}+\hat \rho_\text{stat}^{(1)})\bigr] + 
		\text{Tr}\bigl[\mathbb{J}^{(2)} e^{\mathbb{W}^{(1)}_{\! 0} \tau} \mathcal{J}^{(1)} \hat \rho^{(0)}_{\text{stat}}  \bigr] + \text{Tr}\bigl[ \mathcal{J}^{(1)} e^{\mathbb{W}^{(1)}_{\! 0} \tau} \mathbb{J}^{(2)} \hat \rho^{(0)}_{\text{stat}}\bigr]\\
		& + \Gamma \,\text{Tr}\bigl[ \mathcal{J}^{(1)} \, \partial_ \Gamma[e^{(\mathbb{W}^{(1)}_{\! 0}\!+ \mathcal{L}^{(2)}_0) \tau}]_{\Gamma=0}\, \mathcal{J}^{(1)} \hat \rho_\text{stat}^{(0)} \bigr] +\, \text{Tr}\bigr[\mathcal{J}^{(1)}e^{\mathbb{W}^{(1)}_{\! 0} \tau} \mathbb{W}^{(1)}_{\! 0} \partial_z\widetilde{\mathcal{J}}^{(1)} \hat \rho^{(0)}_{\text{stat}} \bigl]\, , 
	\end{aligned}
\end{equation}
with
\begin{equation}\label{eq:normleadnext}
	\frac{1}{\braket{\tau}_2}=\,\text{Tr}\left[\mathcal{J}^{(1)}(\hat \rho_\text{stat}^{(0)}+\hat \rho_\text{stat}^{(1)}) + \mathbb{J}^{(2)}\hat \rho_\text{stat}^{(0)} \right]\\
	- \text{Tr}\left[\mathcal{J}^{(1)} \partial_z\widetilde{\mathcal{J}}^{(1)} \hat \rho_\text{stat}^{(0)} \right]\, .
\end{equation}
The derivative $\partial_ \Gamma[e^{(\mathbb{W}^{(1)}_{\! 0}\!+ \mathcal{L}^{(2)}_0) \tau}]_{\Gamma=0}$ includes only powers of $\Gamma \tau$. In the long-time limit, one may replace it by $e^{(\mathbb{W}^{(1)}_{\! 0}\!+ \mathcal{L}^{(2)}_0) \tau}$ in order to approximate the slope of the distribution more accurately.
The stationary density matrix in leading and next-to-leading order is determined by $(\mathbb{W}_{\! 1}^{(1)} + \mathbb{W}_{\! 1}^{(2)})\hat \rho_\text{stat}=0$ with $\text{Tr}[\hat \rho_\text{stat}]=1$. Then, $\hat \rho_\text{stat}^{(0)}=[\hat \rho_\text{stat}]_{\Gamma=0}$ and $\hat \rho_\text{stat}^{(1)}=\partial_\Gamma [\hat \rho_\text{stat}]_{\Gamma=0}$. Non-Markovian terms include the operators $\partial_z \mathbb{W}_{\! s}^{(1)}$ or $\partial_z \widetilde{\mathbb{W}}_{\! s}^{(1)}$.
They are neglected under Markov approximation.

\section{Memory and renormalization effects}

\begin{figure}[t]
	\includegraphics[width=0.5\textwidth]{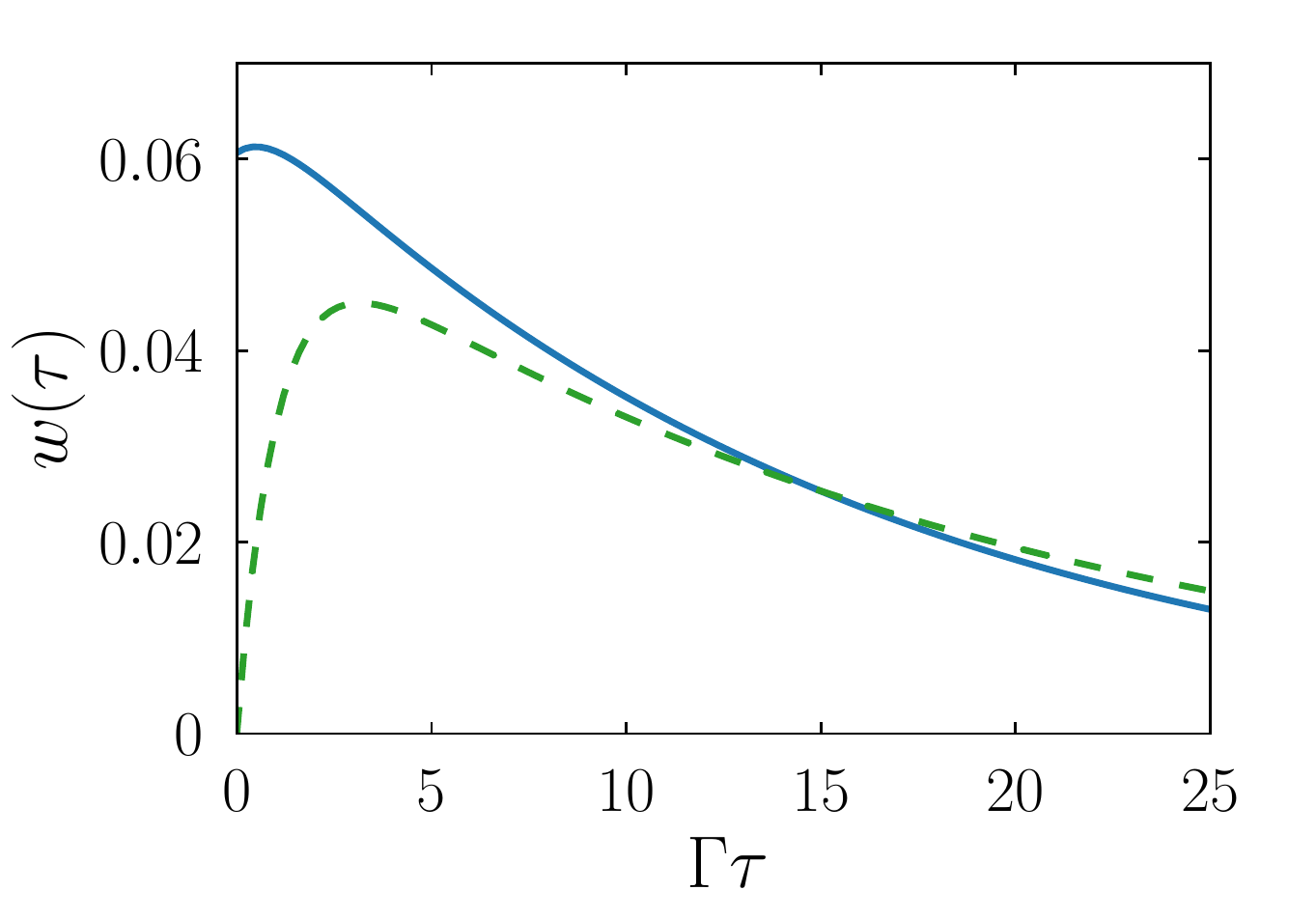}
	\caption{Distribution of electron waiting times for the setup depicted in Fig.~1 of the main text. The approximation of lowest-order sequential tunneling given in \eq{wle} (dashed green line) is compared with the full expression given in \eq{wleadnext} evaluated up to second order in the tunnel-coupling~$\Gamma$ (solid blue line). The level position is $\varepsilon = 0.3\,U$ and no magnetic field is applied. The temperature and the coupling are $k_\tx{B}T=\,U/12$ and $\hbar\Gamma=U/60$.}\label{fig:fig3}
\end{figure}

Our diagrammatic approach yields analytic expressions for the waiting time distribution giving us additional insight into the underlying tunneling dynamics. We obtain for the full waiting time distribution (\ref{eq:wleadnext}) in Fig.~\ref{fig:fig3} (solid blue line) at $\tau = 0$
\begin{equation}
\frac{w_2(0)}{\braket{\tau}_2}= \sum_{\sigma \sigma'}{\rho^{(0)}_\tx{stat}}\hspace{0.0cm}^{\sigma'}_{\sigma'} \Gamma^{\tx{sq}}_{\sigma' 0}  \left[ \Gamma^{\tx{sq}}_{0 \sigma} (\partial_z \Gamma^{\tx{sq}}_{\sigma 0})  + (\partial_z \Gamma^{\tx{sq}}_{0 \sigma}) \Gamma^{\tx{sq}}_{\sigma 0} \right]  
 -  \sum_{\sigma} {\rho^{(0)}_\tx{stat}}\hspace{0.0cm}^{0}_{0} (\partial_z \Gamma^{\tx{vir}}_{00}) \Gamma^{\tx{sq}}_{0 \sigma} \Gamma^{\tx{sq}}_{\sigma 0}\, ,
\end{equation}
with $\Gamma^\tx{sq}_{\chi' \chi}={\mathbb{W}}_{11}^{(1)}\hspace{0.0cm}^{\chi' \chi}_{\chi' \chi}$ and $\Gamma^{\tx{vir}}_{\chi \chi}={\mathbb{W}}_{11}^{(1)}\hspace{0.0cm}^{\chi \chi}_{\chi \chi}$.
Derivatives $\partial_z$ indicate the non-Markovian nature of the process. The first (second) term accounts for memory generated after (before) counting starts, i.e., the second (third) term in \eq{masternfin}. Noteworthy, non-Markovian correction due to virtual charge fluctuations ($\partial_z \Gamma^{\tx{vir}}_{00}$) enter via the second term.

The lifting of the short-time suppression in Fig.~1 of the main text is due to second-oder Markovian corrections. We obtain
\begin{equation} 
\frac{w_2(0)}{\braket{\tau}_2}={\rho^{(1)}_\tx{stat}}\hspace{0.0cm}^{\tx{d}}_{\tx{d}} \Gamma^{\tx{sq}}_{\tx{d}\uparrow}\Gamma^{\tx{sq}}_{\uparrow0}\,,
\end{equation}
at $\tau=0$ with the density-matrix element ${\rho^{(1)}_\tx{stat}}\hspace{0.0cm}^{\tx{d}}_{\tx{d}} = {\Gamma^{\text{sq}}_\tx{ren\,}}_{\downarrow \tx{d}}/(2\Gamma)$. The term ${\Gamma^{\text{sq}}_\tx{ren\,}}_{\downarrow \tx{d}}$ is a correction due to a renormalized tunnel-coupling strength, see \Sec{kernelfig3}.
The distribution at $\tau=0$ becomes most pronounced for $\Delta \approx U$ when the excitation energies $E_\text{d}-E_\uparrow$ and $E_\uparrow-E_0$ align.
The suppression is lifted if $\ket{\downarrow}$ is the ground state of the dot (i.e., $-\frac{\Delta}{2}-U<\varepsilon<\frac{\Delta}{2}$). Then, an additional electron enters frequently due to a second-order Markovian tunneling event. Next, a spin-$\downarrow$ electron tunnels out (i.e., $\frac{\Delta}{2} -U < \varepsilon$) followed by a spin-$\uparrow$ electron (i.e., $-\frac{\Delta}{2} < \varepsilon$). In summary, the lifting can be observed for $\max(\frac{\Delta}{2}-U, -\frac{\Delta}{2} )< \varepsilon < \frac{\Delta}{2}$ and, therefore, a nonvanishing Zeeman splitting is required. The criterion is, especially, not fulfilled in Fig.~\ref{fig:fig3}.